\documentclass[12pt]{article}

\usepackage{bm}

\begin{document}

\title{{\bf Beginnings of the Helicity Basis in the $(S,0)\oplus (0,S)$ Representations of the Lorentz Group}}

\author{{\bf Valeriy V. Dvoeglazov}\\ UAF, Universidad Aut\'onoma de Zacatecas, M\'exico\\E-mail: valeri@fisica.uaz.edu.mx}

\date{\empty}

\maketitle

\begin{abstract}
We write solutions of relativistic quantum equations explicitly in the helicity basis for $S=1/2$ and $S=1$.
We present the analyses of relations between Dirac-like and Majorana-like field operators.
Several interesting features of bradyonic and tachyonic solutions are presented.
\end{abstract}

\section{Introduction.}

In Refs.~\cite{DV1,DV2} we considered the procedure of construction of the field operators {\it ab initio} 
(including for neutral particles). The Bogoliubov-Shirkov method has been used. 
In the present article we investigate the helicity $h=1/2$ and $h=1$ cases in the helicity basis. 
We look for relations between the Dirac-like field operator and the Majorana-like field operator. 

In the first part we refer to the previously found contradiction in the construction of the Majorana-like field operator
for spin 1/2. In the 2nd part we analize the Majorana-like field operator in the $(1,0)\oplus (0,1)$ representation.
It seems that the calculations in the helicity basis only give mathematically and physically reasonable results.

\newpage

\section{Helicity Basis in the $(1/2,0)+(0,1/2)$ Representation.}

The Dirac equation is:
\begin{equation}
[i\gamma^\mu \partial_\mu -mc/\hbar]\Psi (x) =0\,,\quad \mu =0,1,2,3\,.\label{Dirac}
\end{equation}
The $\gamma^\mu$ are the Clifford algebra matrices 
\begin{equation}
\gamma^\mu \gamma^\nu +\gamma^\nu \gamma^\mu = 2g^{\mu\nu}\,,
\end{equation}
$g^{\mu\nu}$ is the metric tensor.
Usually, everybody uses the definition of the field operator (in Ref.~\cite{Itzyk}) in the pseudo-Euclidean metrics
as given {\it ab initio}.
After actions of the Dirac operator on  $\exp (\mp ip_\mu x^\mu)$ the 4-spinors ( $u-$ and $v-$ ) 
satisfy the momentum-space equations: $(\hat p - m) u_h (p) =0$ and $(\hat p + m) v_h (p) =0$, respectively; the $h$ is 
the polarization index. It is easy to prove from the characteristic equations
$Det (\hat p \mp m) =(p_0^2 -{\bf p}^2 -m^2)^2= 0$ that the solutions should satisfy the energy-momentum relations $p_0= \pm E_p =\pm \sqrt{{\bf p}^2 +m^2}$ for both $u-$ and $v-$ solutions.

However, the general scheme of construction of the field operator has been presented in~\cite{Bogoliubov}. In the case of
the $(1/2,0)\oplus (0,1/2)$ representation we have:
\begin{eqnarray}
&&\Psi (x) = {1\over (2\pi)^3} \int d^4 p \,\delta (p^2 -m^2) e^{-ip\cdot x}
\Psi (p) =\nonumber\\
&=& {1\over (2\pi)^3} \sum_{h}^{}\int d^4 p \, \delta (p_0^2 -E_p^2) e^{-ip\cdot x}
u_h (p_0, {\bf p}) a_h (p_0, {\bf p}) =\nonumber\\
&=&{1\over (2\pi)^3} \int {d^4 p \over 2E_p} [\delta (p_0 -E_p) +\delta (p_0 +E_p) ] 
[\theta (p_0) +\theta (-p_0) ] e^{-ip\cdot x} \times \nonumber\\
&\times&\sum_{h}^{} u_h (p) a_h (p)= {1\over (2\pi)^3} \sum_h^{} \int {d^4 p \over 2E_p}
[\delta (p_0 -E_p) +\delta (p_0 +E_p) ] \times\label{fo}\\
&\times&\left [\theta (p_0) u_h (p) a_h (p) e^{-ip\cdot x}  +
\theta (p_0) u_h (-p) a_h (-p) e^{+ip\cdot x} \right ] =\nonumber\\
&=& {1\over (2\pi)^3} \sum_h^{} \int {d^3 {\bf p} \over 2E_p} \theta(p_0)  
\left [ u_h (p) a_h (p)\vert_{p_0=E_p} e^{-i(E_p t-{\bf p}\cdot {\bf x})}+ \right. \nonumber\\
&+& \left. u_h (-p) a_h (-p)\vert_{p_0=E_p} e^{+i (E_p t- {\bf p}\cdot {\bf x})} 
\right ]\nonumber
\end{eqnarray}
During the calculations we had to represent $1=\theta (p_0) +\theta (-p_0)$ above
in order to get positive- and negative-frequency parts. Moreover, we did not yet assumed, which equation this
field operator  (namely, the $u-$ spinor) satisfies, with negative- or positive- mass and/or $p^0= \pm E_p$.

In general we should transform $u_h (-p)$ to the $v_h (p)$. The procedure is given below~\cite{DV1,DV2}.

The explicit forms of the 4-spinors are very well known in the spinorial basis:
\begin{eqnarray}
u_\sigma ({\bf p})= \frac{N_\sigma^+}{2\sqrt{m (E_p +m)}}\pmatrix{[E_p+m+{\bm \sigma}\cdot {\bf p}]\phi_\sigma ({\bf 0})\cr
[E_p+m-{\bm \sigma}\cdot {\bf p}]\chi_\sigma ({\bf 0})\cr}\,,\quad v_\sigma ({\bf p}) =\gamma^5  u_\sigma ({\bf p})\,,\label{s1a}
\end{eqnarray}
where $\phi_\uparrow ({\bf 0}) =\chi_\uparrow  ({\bf 0})= \pmatrix{1\cr 0\cr}$ 
and $\phi_\downarrow ({\bf 0}) =\chi_\downarrow  ({\bf 0})= \pmatrix{0\cr 1\cr}$. The transformation to  
the standard basis is produced with 
the $(\gamma^5 +\gamma^0)/\sqrt{2}$ matrix. The normalizations, projection operators, 
propagators, dynamical invariants {\it etc} have been given in~\cite{Ryder}, for example.

We should assume the following relation in the field operator (\ref{fo}):
\begin{equation}
\sum_{h=\uparrow\downarrow}^{} v_h (p) b_h^\dagger (p) = \sum_{h=\uparrow\downarrow}^{} u_h (-p) a_h (-p)\,.\label{dcop}
\end{equation}
We need $\Lambda_{\mu\lambda} (p) = \bar v_\mu (p) u_\lambda (-p)$.
In the spinorial basis by direct calculations,  we find
$\Lambda_{\mu\lambda} = - im ({\bm \sigma}\cdot {\bf n})_{\mu\lambda}$, ${\bf n} = {\bf p}/\vert{\bf p}\vert$,
provided that the normalization was chosen to the mass $m$. The indices $h$ and $\sigma,\mu, \lambda$ are the corresponding polarization indices.
However, in the helicity basis with the helicity operator 
\begin{eqnarray}
{1\over 2} {\bm \sigma}\cdot\widehat
{\bf p} = {1\over 2} \pmatrix{\cos\theta & \sin\theta e^{-i\varphi}\cr
\sin\theta e^{+i\varphi} & - \cos\theta\cr}\,
\end{eqnarray}
the 2-eigenspinors can be defined as follows~\cite{Var,DVOF}:
\begin{eqnarray}
\phi_{{1\over 2}\uparrow}=\pmatrix{\cos{\theta \over 2} e^{-i\varphi/2}\cr
\sin{\theta \over 2} e^{+i\varphi/2}\cr}\,,\quad
\phi_{{1\over 2}\downarrow}=\pmatrix{\sin{\theta \over 2} e^{-i\varphi/2}\cr
-\cos{\theta \over 2} e^{+i\varphi/2}\cr}\,,\quad\label{ds}
\end{eqnarray}
for $\pm 1/2$ eigenvalues, respectively.

We can start from the Klein-Gordon equation, generalized for
describing the spin-1/2  particles (i.~e., two degrees
of freedom); $c=\hbar=1$:
\begin{equation}
(p_0+{\bm \sigma}\cdot {\bf p}) (p_0- {\bm \sigma}\cdot {\bf p}) \phi
= m^2 \phi\,.\label{de}
\end{equation}
If the $\phi_{\uparrow\downarrow}$ spinors are defined by the equation (\ref{ds}) then we
can construct the corresponding $u-$ and $v-$ 4-spinors:
\begin{eqnarray}
u_\uparrow ({\bf p}) &=&
N_\uparrow^+ \pmatrix{\phi_\uparrow\cr {E_p-p\over m}\phi_\uparrow\cr} =
{1\over \sqrt{2}}\pmatrix{\sqrt{{E_p+p\over m}} \phi_\uparrow\cr
\sqrt{{m\over E_p+p}} \phi_\uparrow\cr}\,,\nonumber\\
u_\downarrow ({\bf p}) &=& N_\downarrow^+ \pmatrix{\phi_\downarrow\cr
{E_p+p\over m}\phi_\downarrow\cr} = {1\over
\sqrt{2}}\pmatrix{\sqrt{{m\over E_p+p}} \phi_\downarrow\cr \sqrt{{E_p+p\over
m}} \phi_\downarrow\cr}\,,\label{s1}\\
v_\uparrow ({\bf p}) &=& N_\uparrow^- \pmatrix{\phi_\uparrow\cr
-{E_p-p\over m}\phi_\uparrow\cr} = {1\over \sqrt{2}}\pmatrix{\sqrt{{E_p+p\over
m}} \phi_\uparrow\cr
-\sqrt{{m\over E_p+p}} \phi_\uparrow\cr}\,, \nonumber\\
v_\downarrow ({\bf p}) &=& N_\downarrow^- \pmatrix{\phi_\downarrow\cr
-{E_p+p\over m}\phi_\downarrow\cr} = {1\over
\sqrt{2}}\pmatrix{\sqrt{{m\over E_p+p}} \phi_\downarrow\cr -\sqrt{{E_p+p\over
m}} \phi_\downarrow\cr}\,,\label{s2}
\end{eqnarray}
where the normalization to the unit
($\pm 1$) was now used:
\begin{eqnarray}
\bar u_h ({\bf p}) u_{h^\prime} ({\bf p}) &=& \delta_{hh^\prime}\,,
\bar v_h ({\bf p}) v_{h^\prime} ({\bf p}) = -\delta_{hh^\prime}\,,\\
\bar u_h ({\bf p}) v_{h^\prime} ({\bf p}) &=& 0 =
\bar v_h ({\bf p}) u_{h^\prime} ({\bf p})\,.
\end{eqnarray}
The commutation
relations may be assumed to be the standard
ones~\cite{Bogoliubov,Itzyk,Wein,Greinb}
(compare  with~\cite{Tokuoka})
\begin{eqnarray}
\vspace{-15mm}&&\left [a_h ({\bf p}),
a_{h^\prime}^\dagger ({\bf k})
\right ]_+ = 2E_p \delta^{(3)} ({\bf p}-{\bf k})
\delta_{hh^\prime}\,,
\left [a_h ({\bf p}),
a_{h^\prime} ({\bf k})\right ]_+ = 0 =
\left [a_h^\dagger ({\bf p}),
a_{h^\prime}^\dagger ({\bf k})\right ]_+\nonumber\\
\vspace{-15mm}&&\\
\vspace{-15mm}&&\left [a_h ({\bf p}),
b_{h^\prime}^\dagger ({\bf k})\right ]_+ = 0 =
\left [b_h ({\bf p}),
a_{h^\prime}^\dagger ({\bf k})\right ]_+,\\
\vspace{-15mm}&&\left [b_h ({\bf p}),
b_{h^\prime}^\dagger ({\bf k})
\right ]_+ = 2E_p \delta^{(3)} ({\bf p}-{\bf k})
\delta_{hh^\prime}\,,
\left [b_h ({\bf p}),
b_{h^\prime} ({\bf k})\right ]_+ = 0 =
\left [b_h^\dagger ({\bf p}),
b_{h^\prime}^\dagger ({\bf k})\right ]_+\nonumber\\
\vspace{-15mm}&&
\end{eqnarray}
Other details of the helicity basis are given in Refs.~\cite{BLP,GR,DVIJTP}.
However, in this helicity case we construct
\begin{equation}
\Lambda_{\mu\lambda} (p) = \bar v_\mu (p) u_\lambda (-p)=
i\sigma^y_{\mu\lambda}\,.\label{LambdaHB}
\end{equation}

It is well known that  ``{\it particle=antiparticle}" in the Majorana theory~\cite{Majorana}. So, in the language of the quantum field theory we should have 
\begin{equation}
b_\mu (E_p, {\bf p}) = e^{i\varphi} a_\mu (E_p, {\bf p})\,.\label{ma}
\end{equation} 
Usually, different authors use $\varphi = 0, \pm \pi/2$ depending on the metrics and on the forms of the 4-spinors and 
commutation relations, {\it etc.}.
The application of the Majorana anzatz leads to the contradiction in the spinorial basis. Namely, it leads to
existence of the preferred axis in every inertial system (only $p_y$ survives), thus breaking the rotational symmetry of the special relativity.

Next, we can use another Majorana anzatz $\Psi = \pm e^{i\alpha} \Psi^{c}$ with usual definitions
\begin{eqnarray}
{\cal C} =e^{i\vartheta_c}\pmatrix{0&i\Theta\cr -i\Theta & 0\cr} {\cal K}\,,\quad
\Theta = \pmatrix{0&-1\cr 1&0\cr}\,.\label{cco}
\end{eqnarray}
Thus, on using $Cu_\uparrow^\ast ({\bf p}) =-iv_\downarrow ({\bf p})$, $Cu_\downarrow^\ast ({\bf p}) =+iv_\uparrow ({\bf p})$ we come to other relations
between creation/annihilation operators 
\begin{eqnarray}
a_\uparrow^\dagger  ({\bf p}) &=& \pm i e^{-i\alpha} b_\downarrow^\dagger ({\bf p})\,,\\
a_\downarrow^\dagger  ({\bf p}) &=& \mp i e^{-i\alpha} b_\uparrow^\dagger ({\bf p})\,,
\end{eqnarray}
which may be used instead of (\ref{ma}). In the case of $\alpha=\pi/2$ we have similar relations as in (\ref{LambdaHB}), but for positive-energy operators.
Due to the possible signs $\pm$ the number of the corresponding states is the same as in the Dirac case that permits us to have the complete system 
of the Fock states over the $(1/2,0)\oplus (0,1/2)$ representation space in the mathematical sense. Please note that the phase factors may have physical significance in quantum field theories as opposed to the textbook nonrelativistic quantum mechanics, as was discussed recently by several authors.
However, in this case we deal with the self/anti-self charge conjugate quantum field operator instead of the self/anti-self charge conjugate quantum states. Please remember that it is the latter that answer for the neutral particles. The quantum field operator 
contains operators for more than one state, which may be either electrically neutral or charged.

\section{Helicity Basis in the $(1,0)+(0,1)$ Representation.}

The solutions of the Weinberg-like equation
\begin{equation}
[\gamma^{\mu\nu} \partial_\mu \partial_\nu - {(i\partial/\partial t)\over E} 
m^2 ] \Psi (x) =0\,.
\end{equation}
are found in Refs.~\cite{Sankar,Novozh,DVA,DVO94}. Here they are:
\begin{eqnarray}
&&u_\sigma ({\bf p})= \pmatrix{D^S (\Lambda_R) \xi_\sigma ({\bf 0})\cr 
D^S (\Lambda_L)\xi_\sigma ({\bf 0})\cr}\,,\,\,
v_\sigma ({\bf p})=\pmatrix{D^S (\Lambda_R \Theta_{[1/2]}) \xi_\sigma^\ast ({\bf 0})\cr 
- D^S (\Lambda_L \Theta_{[1/2]})\xi_\sigma^\ast ({\bf 0})\cr}=\Gamma^5 u_\sigma ({\bf p}),\nonumber\\ 
&&\\
&&\Gamma^5 = \pmatrix{1_{3\times 3}&0\cr
0&-1_{3\times 3}\cr}\,,
\end{eqnarray}
in the ``spinorial" representation.
The $D^S$ is the matrix of the $(S,0)$ representation of the spinor group $SL (2,c)$. 

In the $(1,0)\oplus (0,1)$ representation 
the procedure of derivation of the creation operators (in the similar way as in the previous Section) leads to somewhat 
different situation:
\begin{equation}
\sum_{\sigma=0,\pm 1}^{} v_\sigma (p) b_\sigma^\dagger (p) = \sum_{\sigma=0,\pm 1}^{} u_\sigma (-p) a_\sigma (-p)\,.
\end{equation} 
Hence,
\begin{equation}
b_\sigma^\dagger (p) \equiv 0\,.
\end{equation}
However, if we return to the original Weinberg equations $[\gamma^{\mu\nu} \partial_\mu \partial_\nu \pm 
m^2 ] \Psi_{1,2} (x) =0$ with the field operators:
\begin{eqnarray}
\Psi_1 (x) &=& \frac{1}{(2\pi)^3}\sum_\mu \int \frac{d^3 {\bf p}}{2E_p} [ u_\mu ({\bf p}) a_\mu ({\bf p}) e^{-ip\cdot x}
+ u_\mu ({\bf p}) b_\mu^\dagger ({\bf p}) e^{+ip\cdot x}],\\
\Psi_2 (x) &=& \frac{1}{(2\pi)^3}\sum_\mu \int \frac{d^3 {\bf p}}{2E_p} [ v_\mu ({\bf p}) c_\mu ({\bf p}) e^{-ip\cdot x}
+ v_\mu ({\bf p}) d^\dagger_\mu ({\bf p}) e^{+ip\cdot x}],
\end{eqnarray}
we obtain
\begin{eqnarray}
b_\mu^\dagger (p) &=& [1-2({\bf S}\cdot {\bf n})^2]_{\mu\lambda} a_\lambda (-p)\,,\label{ba}\\
d_\mu^\dagger (p) &=& [1-2({\bf S}\cdot {\bf n})^2]_{\mu\lambda} c_\lambda (-p)\,\label{ca}.
\end{eqnarray}
The application of $\overline u_\mu (-p) u_\lambda (-p) = \delta_{\mu\lambda}$ and 
$\overline u_\mu (-p) u_\lambda (p) = [1- 2({\bf S}\cdot {\bf n})^2]_{\mu\lambda}$ prove that the equations
are self-consistent (similarly to the consideration of the $(1/2,0)\oplus (0,1/2)$ representation).
This situation signifies that in order to construct the Sankaranarayanan-Good field operator (which was used by Ahluwalia, Johnson and 
Goldman~\cite{DVA}) we need additional postulates. One can try to construct 
the left- and the right-hand side of the field operator separately each other.
In this case the commutation relations may also be more complicated.

Is it possible to apply the Majorana-like anzatz to the $(1,0)+(0,1)$ fields? It appears that in this basis we also come 
to the same contradictions as before.
We have two equations
\begin{equation}
a_\mu (p) = + e^{-i\varphi} [1-2({\bf S}\cdot {\bf n})^2]_{\mu\lambda} a_\lambda^\dagger (-p)\,,
\end{equation}
and 
\begin{equation}
a_\mu^\dagger (p) = +e^{+i\varphi} [1-2({\bf S}^\ast \cdot {\bf n})^2]_{\mu\lambda} a_\lambda (-p)\,.
\end{equation}
In the basis where $S_z$ is diagonal the matrix $S_y$ is imaginary~\cite{Var}. So, $({\bf S}^\ast \cdot {\bf n})= S_x n_x - S_y n_y + S_z n_z$, and $({\bf S}^\ast\cdot {\bf n})^2 \neq ({\bf S}\cdot {\bf n})^2$ in the case of $S=1$. So, we conclude that there is the same problem in this 
point, in the aplication of the Majorana-like anzatz, as in the case of spin-1/2. Similarly, one can proceed with (\ref{ca}). What we would have in the basis where all ${\bf S}^i_{jk}=-i\epsilon_{ijk}$ are pure imaginary?
Finally, I just want to mention  that the attempts of constructing the self/anti-self charge conjugate states failed in 
Ref.~\cite{DVA1995}. Instead, the $\Gamma^5 S^c_{[1]}-$ self/anti-self conjugate states
have been constructed therein.

Now we turn to the helicity basis. 
The helicity operator in the $(1/2,1/2)$ representation is frequently presented:
\begin{equation}
{({\bf S}^{I}\cdot {\bf p})\over p} = {1\over p} \pmatrix{0&0&0&0\cr 
0&0&-ip^3&ip^2\cr
0&ip^3&0&-ip^1\cr
0&-ip^2&ip^1&0\cr}\,,\,\,
{({\bf S}\cdot {\bf p})\over p} \epsilon^\mu_{\pm 1} = \pm \epsilon^\mu_{\pm 1}\,,\,\,{({\bf S}\cdot {\bf p})\over p} 
\epsilon^\mu_{0,0_t} = 0\,.
\end{equation} 
However, we are aware about some problems with the chosen basis.
The helicity operator is (in the case of ${\bf S}^3$ diagonal):
\begin{eqnarray}
{({\bf S}^{II}\cdot {\bf p})\over p} = {1\over p} \pmatrix{0&0&0&0\cr 
0&p_z&{p_l \over\sqrt{2}}&0\cr
0&{p_r \over\sqrt{2}}&0&{p_l \over\sqrt{2}}\cr
0&0&{p_r \over\sqrt{2}}&-p_z\cr}\,.
\end{eqnarray}
The unitary transformation~\cite[p.55]{Var}
\begin{equation}
U= \pmatrix{0&0&0&0\cr
0&-{1\over\sqrt{2}}&{i\over\sqrt{2}}&0\cr
0&0&0&1\cr
0&+{1\over\sqrt{2}}&{i\over\sqrt{2}}&0\cr}\,,\,\, U{({\bf S}^{I}\cdot {\bf p})\over p} U^\dagger = {({\bf S}^{II}\cdot {\bf p})\over p}
\end{equation}
can be perfomed to transfer operators and polarization vectors from one basis to another.
The first-basis eigenvectors are:
\begin{eqnarray}
&&\epsilon^\mu_{+1}= {1\over \sqrt{2}} {e^{i\alpha}\over p} \pmatrix{0\cr
{-p^1 p^3 +ip^2 p\over \sqrt{(p^1)^2 +(p^2)^2}}\cr {-p^2 p^3 -ip^1
p\over \sqrt{(p^1)^2 +(p^2)^2}}\cr \sqrt{(p^1)^2 +(p^2)^2}\cr}\,,\,
\epsilon^{\mu }_{-1}={1\over \sqrt{2}}{e^{i\beta}\over p}
\pmatrix{0\cr {p^1 p^3 +ip^2 p\over \sqrt{(p^1)^2 +(p^2)^2}}\cr {p^2 p^3
-ip^1 p\over \sqrt{(p^1)^2 +(p^2)^2}}\cr -\sqrt{(p^1)^2 +(p^2)^2}\cr} \nonumber\\
&&\\
&&\epsilon^{\mu }_0={\frac{1}{m}}\pmatrix{p\cr {E \over p}
p^1 \cr {E \over p} p^2 \cr{E \over p} p^3\cr}\,,\quad
\epsilon^{\mu }_{0_{t}}={\frac{1}{m}}\pmatrix{E_p\cr p^1\cr
p^2\cr p^3\cr}\,.
\end{eqnarray}
The eigenvectors $\epsilon^\mu_{\pm  1}$ are not the eigenvectors
of the parity operator ($\gamma_{00} R$) of this representation. However, the $\epsilon^\mu_{1,0}$,
$\epsilon^\mu_{0,0_t}$
are. Surprisingly, the latter have no well-defined massless limit. In order to get the well-known massless limit one should use the basis of the light-front form reprersentation, cf.~\cite{Ahl-lf}. We also note that the polarization vectors have relations to the solutions of the $(1,0)\oplus (0,1)$ representation through the Proca equations or the Duffin-Kemmer-Petiau equations.

The corresponding helicity operator of the $(1,0)\otimes (0,1)$ representation is
\begin{eqnarray}
\hat h =\pmatrix{({\bf S}_{3\times 3}\cdot {\bf p})&0\cr
0&({\bf S}_{3\times 3}\cdot {\bf p})\cr}
\end{eqnarray}
The eigen 3-vectors are~\cite{Var,DVOF}
\begin{eqnarray}
\phi_{\uparrow} &=&
N\,e^{i\vartheta_+}\,\pmatrix{{1\over 2} (1+\cos\theta) e^{-i\varphi}\cr
\sqrt{{1\over 2}} \sin\theta\cr
{1\over 2} (1-\cos\theta) e^{+i\varphi}\cr}\,,\,
\phi_{\downarrow} = N\,e^{i\vartheta_-}\,\pmatrix{-{1\over 2} (1-\cos\theta)
e^{-i\varphi}\cr
\sqrt{{1\over 2}} \sin\theta\cr
-{1\over 2} (1+\cos\theta) e^{+i\varphi}\cr}\nonumber\\
&&\\
&&\qquad\qquad\phi_{\rightarrow} =  N\,e^{i\vartheta_0}\,
\pmatrix{-\sqrt{{1\over 2}}\sin \theta \,e^{-i\varphi}\cr
\cos\theta\cr
\sqrt{{1\over 2}}\sin \theta \,e^{+i\varphi}\cr}\,.
\end{eqnarray}

Finally, some notes concerning with the tachyonic solutions of the Weinberg equations in the $(1,0)\oplus (0,1)$
representation space. While some authors, e.g.~Ref.~\cite{Kapuscik}, argued recently that the tachyonic energy-momentum relation
$E=\pm\sqrt{{\bf p}^2 c^2 - m_0^2 c^4}$ may lead to some interpretational problems, we still consider it in this paper.
The Weinberg equations $[\gamma^{\mu\nu} \partial_\mu \partial_\nu \pm 
m^2 ] \Psi_{1,2} (x) =0$ give us both bradyonic and tachyonic solutions, $E=\pm\sqrt{{\bf p}^2 c^2 \pm m_0^2 c^4}$.
We present them now in the helicity basis which may help us to overcome the difficulties in the construction of the Majoran(-like)
field operators, as shown above. If $\phi_R ({\bf 0}) =\phi_L ({\bf 0})$ the 6-objects can be normalized to the unit.
The solutions of $[\gamma^{\mu\nu} p_\mu p_\nu - 
m^2 ] \Psi (x) =0$ are 
\begin{eqnarray}
u_\uparrow ({\bf p}) &=&
{1\over \sqrt{2}} \pmatrix{{m\over E_p-p} \chi_\uparrow\cr
{E_p -p\over m} \chi_\uparrow\cr}\,,\quad
u_\downarrow ({\bf p}) = {1\over
\sqrt{2}}\pmatrix{{m\over E_p+p} \chi_\downarrow\cr {E_p+p\over
m} \chi_\downarrow\cr}\,,\label{sb1}\\
&&\qquad\qquad u_\rightarrow ({\bf p}) = {1\over
\sqrt{2}}\pmatrix{\chi_\rightarrow \cr \chi_\rightarrow\cr}\,.\label{sb2}
\end{eqnarray}
In the case of tachyonic solutions ($E < p$) we shall be no able to normalize to 1. However, it is possible to normalize to -1.
In this case we have in the helicity basis:
\begin{eqnarray}
U_\uparrow ({\bf p}) &=&
{1\over \sqrt{2}} \pmatrix{{E_p+p\over m} \chi_\uparrow\cr
-{m\over E_p +p} \chi_\uparrow\cr}\,,\quad
U_\downarrow ({\bf p}) = {1\over
\sqrt{2}}\pmatrix{{m\over E_p+p} \chi_\downarrow\cr -{E_p+p\over
m} \chi_\downarrow\cr}\,,\label{st1}\\
&&\qquad\qquad U_\rightarrow ({\bf p}) = {1\over
\sqrt{2}}\pmatrix{\chi_\rightarrow \cr -\chi_\rightarrow\cr}\,.\label{st2}
\end{eqnarray}
Nevertheless, self/anti-self charge-conjugated 6-objects have not been constructed till now.

\section{Conclusions.}

We conclude that the calculations in the helicity basis may be useful to give mathematically and physically reasonable results
when dealing with the Majorana particles.

\medskip

{\bf Acknowledgements.} I acknowledge discussions with colleagues at recent conferences.
I am grateful to the Zacatecas University for professorship.


\smallskip

\begin{thebibliography}{99}

\footnotesize{

\bibitem{DV1} V. V. Dvoeglazov, Hadronic J. Suppl. {\bf 18}, 239 (2003).\\[-6mm]

\bibitem{DV2} V. V. Dvoeglazov, Int. J. Mod. Phys. B{\bf 20}, 1317 (2006).\\[-6mm]

\bibitem{Itzyk} C. Itzykson and J.-B. Zuber, {\it Quantum Field Theory}
(McGraw-Hill Book Co., 1980). Sec. 3-3.\\[-6mm]

\bibitem{Bogoliubov} N. N. Bogoliubov and D. V. Shirkov, {\it Introduction to 
the Theory of Quantized Fields.} 2nd Edition. (Nauka, Moscow, 1973).\\[-6mm]

\bibitem{Ryder} L. H. Ryder, {\it Quantum Field Theory.} (Cambridge University Press, Cambridge, 1985).\\[-6mm]

\bibitem{Var} D. A. Varshalovich, A. N. Moskalev and V. K. Khersonski\u{\i},
{\it Quantum Theory of Angular Momentum} (World Scientific, Singapore, 1988), \S 6.2.5.\\[-6mm]

\bibitem{DVOF} V. V. Dvoeglazov, Fizika B{\bf 6}, 111 (1997).\\[-6mm]

\bibitem{Wein} S. Weinberg, {\it The Quantum Theory of Fields. Vol. I.
Foundations.} (Cambridge University Press, Cambridge, 1995).\\[-6mm]

\bibitem{Greinb} W. Greiner, {\it Field Quantization.} (Springer, 1996).\\[-6mm] 

\bibitem{Tokuoka} Z. Tokuoka, Prog. Theor. Phys. {\bf 37},  603 (1967).\\[-6mm]

\bibitem{GR} H. M. R\"uck and W. Greiner, J. Phys. G: Nucl. Phys. {\bf 3},
657 (1977).\\[-6mm]

\bibitem{DVIJTP} V. V. Dvoeglazov, Int. J. Theor. Phys. {\bf 43}, 1287 (2004).\\[-6mm]

\bibitem{BLP} V. B. Berestetski\u{\i}, E. M. Lifshitz and L. P.  Pitaevski\u{\i}, {\it Quantum
Electrodynamics.} (Pergamon Press, 1982, translated from the Russian), \S
16.\\[-6mm]

\bibitem{Majorana} E. Majorana, Nuovo Cim. {\bf 14}, 171 (1937).\\[-6mm]

\bibitem{Sankar} A. Sankaranarayanan and R. H. Good, Jr., Nuovo Cim. {\bf 36}, 1303 (1965).\\[-6mm]

\bibitem{Novozh} Yu. V. Novozhilov, {\it Introduction to Elementary Particle
Physics} (Pergamon Press, 1975).\\[-6mm]

\bibitem{DVA} D. V. Ahluwalia, M. B. Johnson and T. Goldman, Phys. Lett. B{\bf 316}, 102 (1993).\\[-6mm]

\bibitem{DVO94} V. V. Dvoeglazov, Int. J. Theor. Phys. {\bf 37}, 1915 (1998).\\[-6mm]

\bibitem{DVA1995} D. V. Ahluwalia, Int. J. Mod. Phys. A{\bf 11}, 1855 (1996).\\[-6mm]

\bibitem{Ahl-lf} D. V. Ahluwalia and M. Sawicki, Phys. Rev. D{\bf 47}, 5161 (1993); Phys. Lett. B{\bf 335}, 24 (1994).\\[-6mm]

\bibitem{Kapuscik} E. Kapu\'scik, in  {\it Relativity, Gravitation, Cosmology: Beyond Foundations.}
Ed. by V. V. Dvoeglazov. (Nova Science Pubs., Hauppauge, NY, USA, 2018).\\[-6mm]

}


\end{thebibliography}
\end{document}